\documentclass[aps,prb,preprint,superscriptaddress]{revtex4-1}

\usepackage{graphicx}
\usepackage{color}
\usepackage{amsmath, bm} 
\usepackage[version=3]{mhchem} 
\usepackage[T1]{fontenc}       
\usepackage{siunitx}
\usepackage{physics}
\usepackage{hyperref}
\usepackage{braket}

\newcommand{\VSCZ}{\ce{V_{S}}-\ce{Cu_{Zn}}}

\begin{document}
\title{Lone-pair effect on carrier capture in Cu$_2$ZnSnS$_4$ solar cells} 

\author{Sunghyun Kim} 
\email{sunghyun.kim@imperial.ac.uk}
\affiliation{Department of Materials, Imperial College London, London SW7 2AZ, United
Kingdom}

\author{Ji-Sang Park} 
\affiliation{Department of Materials, Imperial College London, London SW7 2AZ, United
Kingdom}

\author{Samantha N. Hood} 
\affiliation{Department of Materials, Imperial College London, London SW7 2AZ, United
Kingdom}

\author{Aron Walsh} 
\email{a.walsh@imperial.ac.uk}
\affiliation{Department of Materials, Imperial College London, London SW7 2AZ, United
Kingdom}
\affiliation{Department of Materials Science and Engineering, Yonsei University, Seoul 03722, Korea}

\begin{abstract}
The performance of kesterite thin-film solar cells is limited by a low open-circuit voltage due to defect-mediated electron-hole recombination. We calculate the non-radiative carrier-capture cross sections and Shockley-Read-Hall recombination coefficients of deep-level point defects in Cu$_2$ZnSnS$_4$ (CZTS) from first-principles. While the oxidation state of Sn is +4 in stoichiometric CZTS, inert lone pair (5$s^2$) formation lowers the oxidation state to +2. The stability of the lone pair suppresses the ionization of certain point defects, inducing charge transition levels deep in the band gap. We find large lattice distortions associated with the lone-pair defect centers due to the difference in ionic radii between Sn(II) and Sn(IV). The combination of a deep trap level and large lattice distortion facilitates efficient non-radiative carrier capture, with capture cross-sections exceeding $10^{-12}$ cm$^2$. The results highlight a connection between redox active cations and `killer' defect centres that form giant carrier traps. This lone pair effect will be relevant to other emerging photovoltaic materials containing n$s^2$ cations. 
\end{abstract}
\maketitle

\section*{Introduction}
In a semiconductor subject to above-band-gap illumination, the lifetime of charge carriers is determined by the kinetics of electron-hole recombination processes:
radiative, Auger, and trap-assisted recombination. \cite{Nelson:2003is,Park:2018et}
Radiative and Auger recombination usually only become significant at high carrier concentrations
such as in light-emitting diodes or solar cells using concentrated sunlight. 
In most photovoltaic technologies,
defects limit carrier lifetimes and device efficiencies by acting as non-radiative electron-hole recombination centers. \cite{Stoneham:2000jp,Kirchartz:2018hg}

\begin{figure}
\centering
\includegraphics[width=8.3cm]{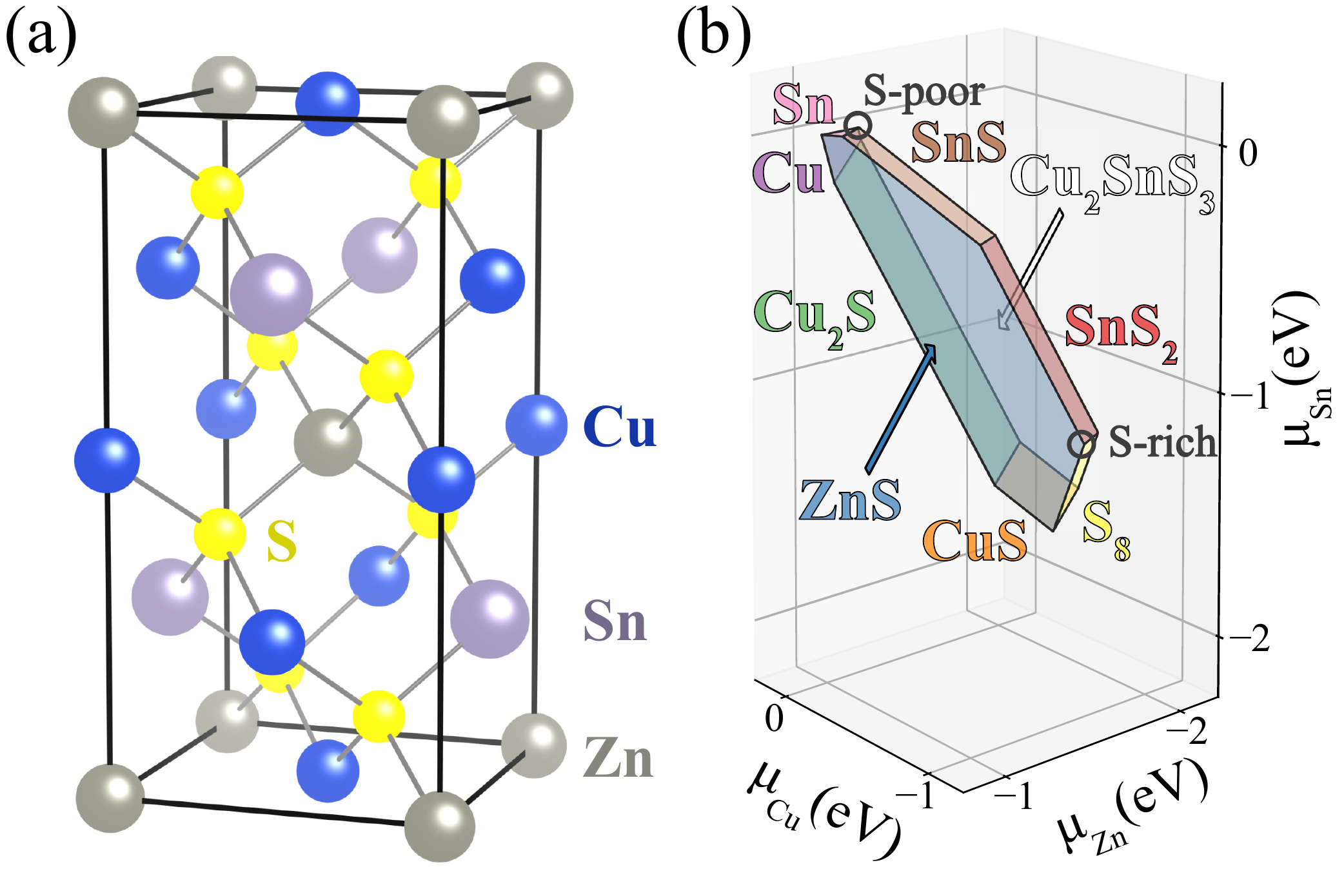}
\caption{\label{fig:struct}
(a) Atomic structure and (b) phase diagram of \ce{CuZnSnS_4} in chemical potential space, 
where $\mu_i = 0$ represent element $i$ in its standard state.
Blue, gray, purple and yellow balls represent Cu, Zn, Sn, and S atoms, respectively.
}
\end{figure}

Thin-film solar cells offer advantages over traditional silicon-based solar cells 
as they require less raw materials and energy to produce, and open up new application areas such as building-integrated photovoltaics.
As an alternative to the current thin-film light absorbers such as CdTe and \ce{Cu(In{,}Ga)Se_2} (CIGS) whose constituting elements are vulnerable to decreases in supply,
kesterite minerals such as \ce{Cu_2ZnSnS_4} (CZTS) and \ce{Cu_2ZnSnSe_4} (CZTSe) (see Fig. \ref{fig:struct} (a)), have attracted much attention due to the earth-abundance of Cu, Zn, and Sn. \cite{Kumar:2013wo,Kumar:2015dd,Kaur:2017eua}
In 2014, an alloy CZTSSe kesterite solar cell reached a record light-to-electricity conversion efficiency of 12.6\%. \cite{Wang:2013gs}
Recently, 11\% efficiency is achieved in a pure sulfide CZTS solar cell. \cite{Yan:2018dw}
However, this technology suffers from a large open-circuit voltage (\ce{V_{OC}}) deficit. \cite{Barkhouse:2011fh,Todorov:2012bo,Grenet:2018ia}
The performance of current kesterite-based solar cells falls far below the Shockley-Queisser limit of $\sim 30\%$. \cite{Shockley:1961co,Ruhle:2016bc}
One likely origin of the \ce{V_{OC}} deficit is a short minority carrier (electron) lifetime of below few ns
due to fast non-radiative recombination pathways. \cite{Wallace:2017hga,Grenet:2018ia}

Thus, it is important to identify dominant recombination centers and to control their concentrations. 
According to Shockley-Read-Hall (SRH) statistics, \cite{Hall:1952iz,Shockley:1952it}
a deep level in the band gap of a semiconductor acts as an efficient recombination channel 
that facilitates the sequential capture of minority and majority carriers.
In addition to deep thermodynamic charge transition levels,
large lattice distortions are required to achieve fast recombination rates. \cite{Stoneham:gb,Stoneham:2000jp}
However, due to the strong interactions between an impurity and a host material,
it is hard to predict the properties (charge transition level and lattice distortion) of impurities \textit{a priori} or to find general trends in various host materials. 
Identification of recombination centers has relied on individual experimental or theoretical studies.
If there exists a simple criterion, 
then we can identify detrimental defects limiting the efficiency more easily and screen candidate optoelectronic materials more efficiently.

In this work, 
we argue that defects in semiconductors involving heavy post-transition metals are likely to act as fast non-radiative recombination centers 
because not only due to their deep nature but also the large lattice distortions
that accompany a change in oxidation state.
As a representative case, we study the native point defects in CZTS containing multivalent ions of Sn and Cu.
Through analysis of carrier capture rates from first-principles, 
we find that the dominant non-radiative recombination centers (\ce{V_S}, \VSCZ{} and \ce{Sn_{Zn}}) are associated with Sn 5$s^2$ lone-pair 
configurations.
They produce deep donor levels due to the Sn double reduction, and
the recombination processes involve large structure distortions
because of the change in the ionic radius of Sn during the carrier capture. 
We expect to find similar behaviour for other lone pair cations including Bi and Sb.

\section*{Methods}

\subsection*{Non-radiative carrier capture}
\begin{figure}[]
\centering
\includegraphics[width=8.3cm]{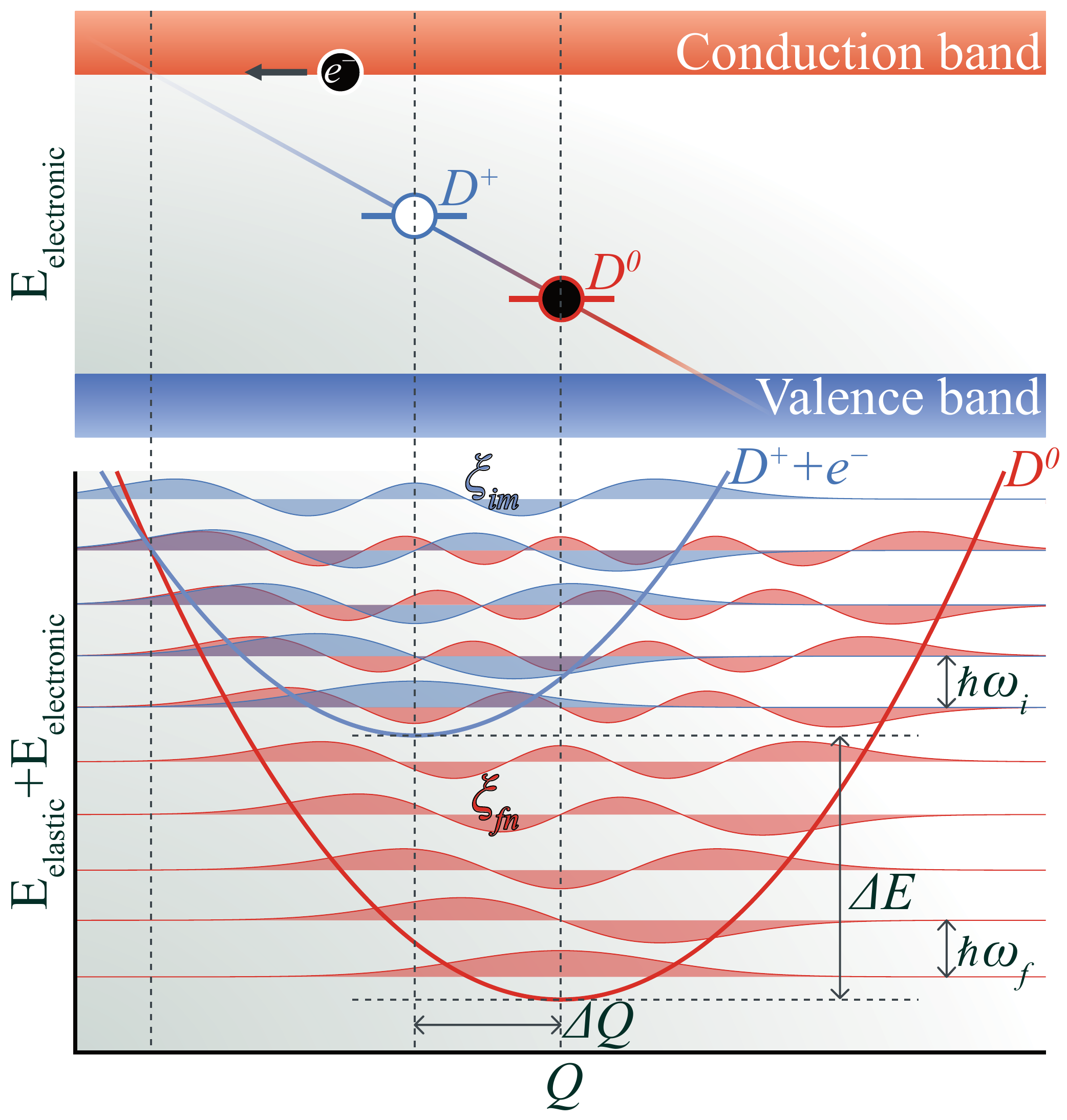}
\caption{\label{fig:nonrad}
Schematic of non-radiative electron ($e^-)$ capture by a positively charged defect ($D^+$) turning into the neutral defect ($D^0$) in band diagram (upper panel) following Henry and Lang (Ref. \citenum{Henry:1977bfa}).
The electron (black circle) is captured by the empty level (blue line) crossing the conduction band due to the thermal vibration of local geometry.
The diagonal line represents evolution of the electronic energy level as the local geometry ($Q$) vibrates.
The occupied level (red line) has a lower electronic energy due to the change in the equilibrium geometry ($\Delta Q$).
Configuration coordinate diagram (lower panel) for the same process shows potential energy surfaces of $D^++e^-$ (blue curve) and $D^0$ (red curve). The vibrational wave functions ($\xi_{im}$ and $\xi_{fn}$, see the text) are shown in lighter shades.
}
\end{figure}

The phenomenon of carrier capture in semiconductors via multiphonon emission has been extensively studied
following pioneering work by Huang and Rhys, \cite{Huang:1950fg} and Henry and Lang. \cite{Henry:1977bfa}
The initial excited state of system, for example, a positively charged donor (\ce{D^+}) with an electron in the conduction band ($e^-$), vibrates around the equilibrium geometry.
Owing to the electron-phonon coupling, the deformation of the structure causes the electronic energy level of a state localized around the defect to oscillate.
As the defect level approaches the conduction band, the probability for the defect to capture an electron increases significantly.
When an electron is captured, the donor becomes neutral (\ce{D^0}) and relaxes to a new equilibrium geometry by emitting multiple phonons as shown in Fig. \ref{fig:nonrad}.
To describe and predict this process 
quantitative accounts of the electronic and atomic structures as well as the vibrational properties of the defect are essential.

Recently, approaches have been developed for first-principles calculations 
of capture rates within a certain set of approximations. \cite{Shi:2012iv,Alkauskas:2014kk}
We have adopted a one-dimensional configuration coordinate for the effective vibrational wave function 
and the static coupling theory for electron-phonon coupling matrix elements as proposed by Alkauskas \textit{et al.}. \cite{Alkauskas:2014kk,Alkauskas:2016kf}

We described the degree of deformation using a one-dimensional configuration coordinate $Q$ defined by 
\begin{equation}
Q^2 = \sum_{\alpha,i} m_{\alpha} \Delta R_{\alpha,i}^2,
\end{equation}
where $m_\alpha$ and $\Delta R_{\alpha, i}$ are the atomic mass and the displacement along the direction $i$ from the equilibrium position of atom $\alpha$, respectively.
The vibrational wave function of excited ($\xi_{im}$) and ground ($\xi_{fn}$) states, and associated frequencies $\omega_{i}$ and $\omega_{f}$ were obtained 
by solving the one-dimensional Schr\"odinger equation for potential energy surfaces around the equilibrium geometries.
The capture coefficient is given by 
\begin{equation}
C = V \frac{2\pi}{\hbar}gW^{2}_{if}\sum_{m}w_m\sum_n |\mel{\xi_{im}}{Q}{\xi_{fn}}|^{2} \delta(\Delta E + m \hbar \omega_{i} - n \hbar \omega_{f}),
\end{equation}
where $V$, $g$ and $W^{2}_{if}$ are the volume of supercell, the degeneracy factor and the electron-phonon coupling matrix element of initial and final states, respectively.
$w_m$ is the occupation number of the excited vibrational state $\xi_{im}$, and $\Delta E$ corresponds to the difference in energy of excited and ground states.

\begin{figure*}[h]
\centering
\includegraphics[width=12.8cm]{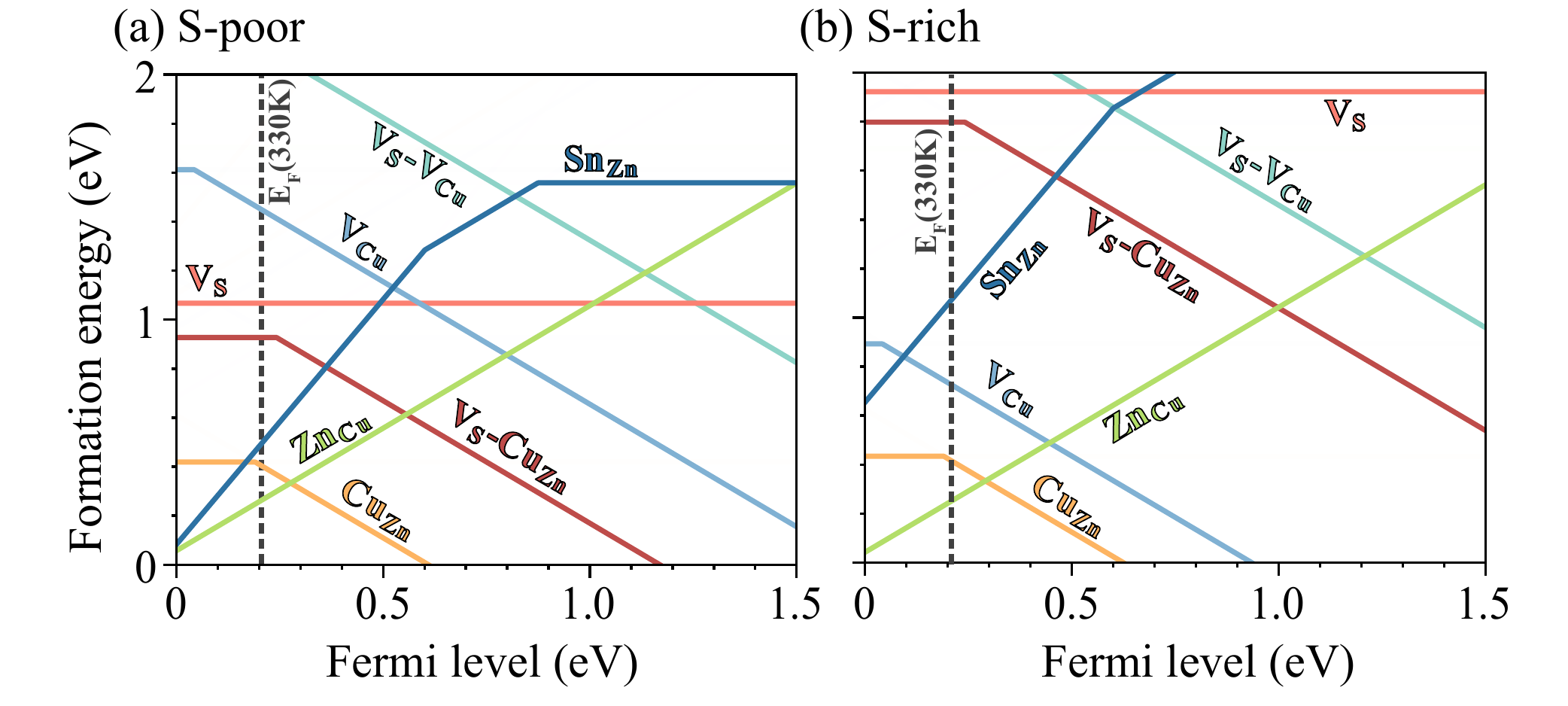}
\caption{\label{fig:form}
Calculated formation energies of native defects in CZTS (a) under S-poor conditions and (b) under S-rich conditions.
The self-consistent Fermi level resulting from the equilibrium defect population is shown as a vertical black dashed line.
The top of valence band is set to 0 eV, while the bottom of conduction band is 1.46 eV.
}
\end{figure*}

The Coulomb interaction at temperature $T$ between a carrier with charge $q$ and a defect in a charge state $Q$ is accounted by the Sommerfeld factor $\langle s \rangle$; \cite{Passler:1976gi,Landsberg:2009uO}
\begin{equation}
\begin{split}
\langle s \rangle=\left\{
        \begin{array}{llll}
            4 |Z|(\pi E_R / k_B T)^{\frac{1}{2}},  \mathrm{for}~ Z<0\\\\
            8/\sqrt{3} (\pi^2 Z^2E_R/k_BT)^{\frac{2}{3}} \\
            \times\mathrm{exp}(-3(Z^2\pi^2E_R/k_B T)^{\frac{1}{3}})
            , \mathrm{for}~ Z>0,
        \end{array}
    \right.
\end{split} 
\end{equation}
where $k_B$ is the Boltzmann constant.
$E_R=m^*q^4/(2\hbar^2\varepsilon^2)$ is an effective Rydberg energy where $m^*$ and $\varepsilon$ are an effective mass of the carrier and a low-frequency dielectric constant, respectively.
For an attractive center, $Z=Q/q$ is negative, while $Z$ is positive for a repulsive center.

Based on the principle of detailed balance, the steady-state recombination rate $R$ 
via a defect with electron capture cross section $\sigma_n$ and hole capture cross section $\sigma_p$
is given by \cite{Shockley:1952it,Hall:1952iz}
\begin{equation}
R=\frac{np-n^2_{i}}{\tau_{p}(n+n_{1})+\tau_{n}(p+p_{1})},
\end{equation}
where 
\begin{equation}
\begin{split}
\tau^{-1}_{n}=N_{t}\sigma_n v_{th,n}=N_{t}C_{n}, \\
\tau^{-1}_{p}=N_{t}\sigma_p v_{th,p}=N_{t}C_{p}.
\end{split}
\end{equation}
Here, $n$, $p$ and $N_{t}$ denote concentrations of electrons, holes and defects, respectively.
$n_{1}$ and $p_{1}$ represent the densities of electrons and holes, respectively, when the Fermi level is located at the trap level.
The thermal velocities of electrons $v_{th,n}$ and holes $v_{th,p}$ are calculated from the effective masses
in the electronic band structure.
$C_{n}$ and $C_{p}$ are the capture coefficients for electron and hole, respectively.

\subsection*{Electronic structure theory}
The atomic and electronic structure of defects were calculated from 
first-principles  within the framework of density functional theory (DFT) \cite{Hohenberg1964,Kohn1965}.
We employed the projector-augmented wave (PAW) method \cite{Blochl1994} 
and the hybrid exchange-correlation functional of Heyd-Scuseria-Ernzerhof (HSE06) \cite{Heyd2003},
as implemented in VASP \cite{Kresse1999}.
The wave functions were expanded in plane waves up to an energy cutoff of 380 eV. 
A Monkhorst-Pack \textit{k}-mesh \cite{Monkhorst1976} with a grid spacing less than 2$\pi\times$0.03 {\AA}$^{-1}$ was used for Brillouin zone integration.
The atomic coordinates were optimized until the residual forces were less than 0.02 eV/\AA.
The lattice vectors were relaxed until stress was below 0.5 kbar.
For defect formation, a $2 \times 2 \times 1$ supercell expansion (64 atoms) of the conventional cell was employed.

We calculated the formation energy $\Delta E_{form}(D^q)$ of a defect $D$ in the charge state $q$ which is given by \cite{Freysoldt:2014ej}
\begin{equation}
\Delta E_{form}(D^q) = E_{tot}(D^q) - E_{tot}(bulk) - \sum_i n_i \mu_i + qE_F + E_{corr},
\end{equation}
where $E_{tot}(bulk)$ and $E_{tot}(D^q)$ are the total energies of a bulk supercell and a supercell containing the defect $D^q$, respectively.
In the third term on the right-hand side, 
$\mu_i$ and $n_i$ are the chemical potential and number of atoms $i$ added to the supercell, respectively.
$E_F$ is the Fermi level, and $E_{corr}$ is a correction term to 
account for the  artificial electrostatic interaction due to periodic boundary conditions. \cite{Freysoldt:2009ih,Kumagai:2014ih}
The formation energy is a function of the Fermi level, 
while the Fermi level is determined by the concentration of defects.
Thus, we calculated the equilibrium concentration of defects and the Fermi level self-consistently, 
under the constraint of charge neutrality condition for overall system of defects and charge carriers, using the SC-FERMI code. \cite{scfermi}

\section*{Results and discussion}

\subsection*{Equilibrium phase diagram}
A challenge to achieving high efficiency from kesterite thin-film solar cell is
to synthesize homogeneous CZTS without unintentional formation of secondary phases. \cite{Scragg:2008kw, Chen:2010cz, Jackson:2014dy, Dimitrievska:2016cj, Wallace:2017hga}
The thermodynamic chemical potential $\mu$ of each element depends 
on the growth environment including partial pressures and temperature.
We compare the DFT total energies of CZTS and its competing phases in the chemical potential space (Fig. \ref{fig:struct} (b)),
showing the range of chemical potentials that favors the formation of CZTS,
using CPLAP. \cite{Buckeridge:2014ht}
The narrow range and complex shape of the phase diagram implies that
it is hard to get a single-phase and homogeneous CZTS sample without the secondary phases.
Even `pure' CZTS is expected to contain an equilibrium population of point defects whose concentrations are controlled by the chemical potentials.
We calculate the formation energies of the native defects under S-poor and S-rich conditions depicted in the phase diagram (Fig. \ref{fig:struct} (b)).

\begin{figure*}
\centering
\includegraphics[width=12.8cm]{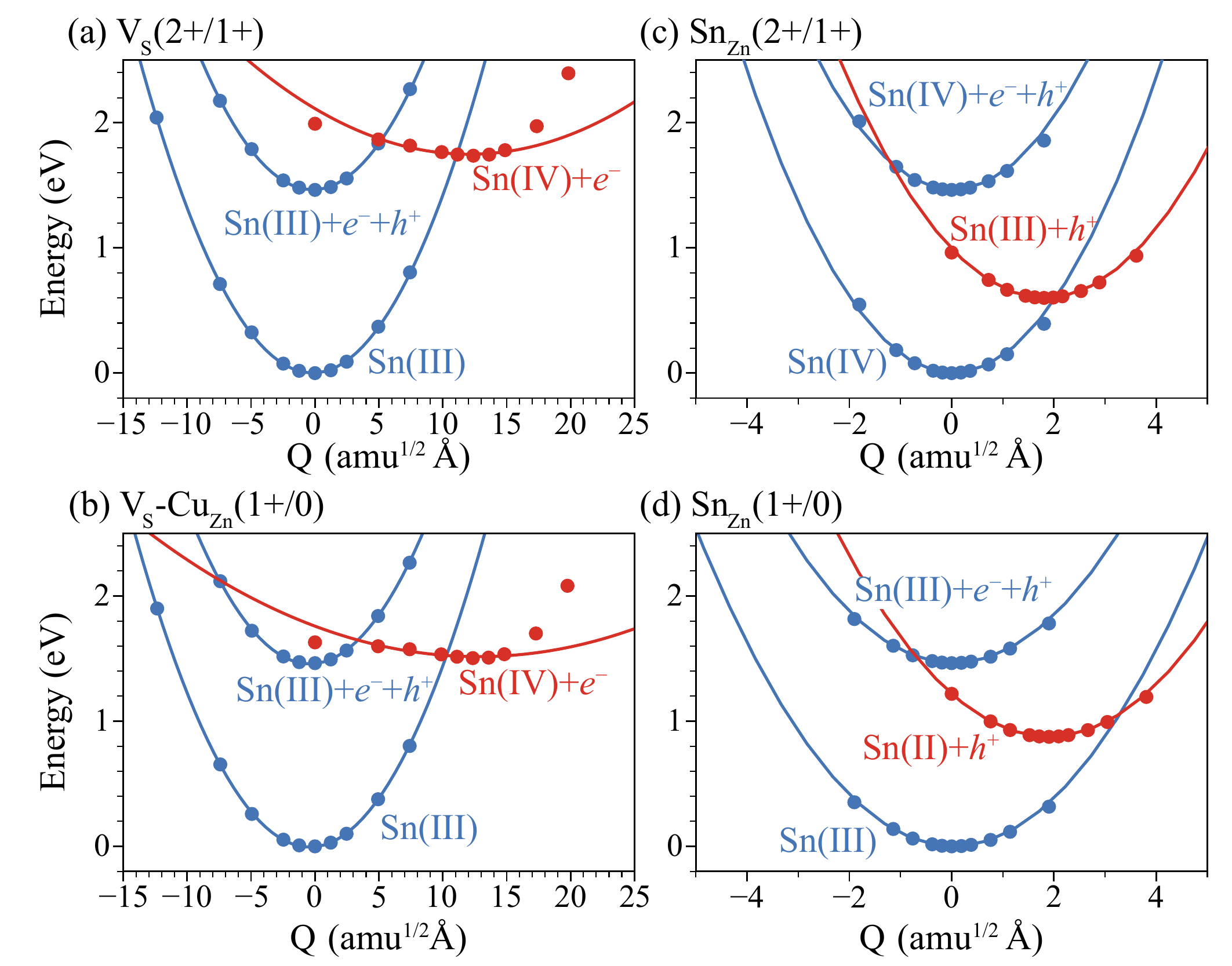}
\caption{\label{fig:cc}
Configuration coordinate diagrams for 
(a) \ce{V_S} (2+/1+), (b) \VSCZ{} (0/1+), (c) \ce{Sn_{Zn}} (2+/1+) and (c) \ce{Sn_{Zn}} (1+/0).
The dot represents the formation energy calculated by DFT,
and the line is a quadratic fit to the change in energy as the structure is distorted along the configuration coordinate. 
$Q$ defines a pathway between the minimum-energy structure for each charge state.
}
\end{figure*}

\subsection*{S-poor growth environment}
Under S-poor conditions, which could be realized by annealing in a low sulfur partial pressure, the most dominant native defects are \ce{Cu_{Zn}} and \ce{Zn_{Cu}} antisites 
which are shallow and responsible for the p-type behaviour with a Fermi level 
close to the valence band
(see Fig. \ref{fig:form} (a)).
At the Fermi level of 0.22 eV determined self-consistently,
we predict high concentrations of \ce{V_S} ($\num{1.3E16}$ \ce{cm^{-3}}), \VSCZ{} ($\num{3.0E17}$ \ce{cm^{-3}}) and \ce{Sn_{Zn}} ($\num{1.1E18}$ \ce{cm^{-3}}).
Here, we assume the growth and annealing temperature of 853 K resulting in defect populations and the operating temperature of 330 K to equilibrate the Fermi level.

Previously, we have shown that \ce{V_S} can act as an efficient non-radiative recombination center in CZTS. \cite{Kim:2018jd}
However, for electron capture, \ce{V_S} needed to be activated.
Firstly, as the ground state of \ce{V_S^0} involving Sn(II) is neutral and produces a state resonant within the valence band,
thermal excitation is required to access  \ce{V_S^+}.
As shown in Fig. \ref{fig:cc} (a), the hole capture barrier for \ce{V_S^+} is so high that the thermal motion can not overcome it. 
Instead, the optical absorption can trigger the vertical transition from \ce{V_S^+} to \ce{V_S^{2+}}, which corresponds to 
Sn(III) to Sn(IV) oxidation.

\begin{figure*}[]
\centering
\includegraphics[width=12.8cm]{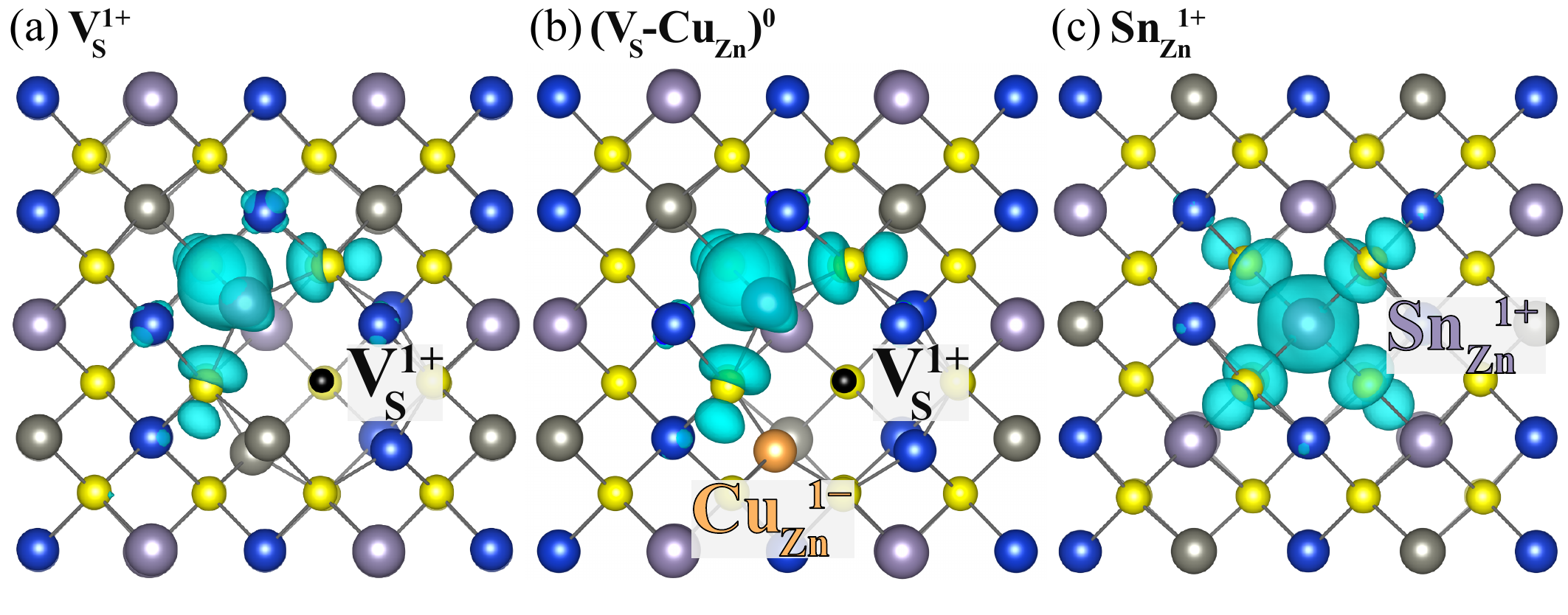}
\caption{\label{fig:charge}
Charge densities of the lowest unoccupied Kohn-Sham orbitals associated with 
(a) \ce{V^{1+}_S}, (b) neutral (\ce{V_S}-\ce{V_{Cu}})$^0$ and (c) \ce{Sn^1+_{Zn}}.
The black and orange balls represent \ce{V_S} and \ce{Cu_{Zn}}, respectively.
}
\end{figure*}

Here, we find that \ce{V_S} can also be activated by forming a defect complex with \ce{Cu_{Zn}}.
In (\VSCZ)$^0$, the electronic wave function is localized around the
Sn 5$s$ lone-pair orbital similar to that of \ce{V^{1+}_{S}} (Fig. \ref{fig:charge} (a) and (b)),
suggesting that the ionized acceptor \ce{Cu^{1-}_{Zn}} ionizes the neutral donor \ce{V^{0}_S}.
Thus, Sn(III) becomes the ground-state electronic configuration in the neutral \VSCZ{} complex,
indicating, unlike the isolated \ce{V_S}, thermal excitation is not necessary.

We further find that optical excitation is not required for hole capture by the \VSCZ{} complex.
As a stronger Coulomb force binds the negatively charged acceptor \ce{Cu^{1-}_{Zn}},
the formation energy difference between Sn(III) and Sn(IV) is reduced in the \ce{V_S}-\ce{V_{Cu}} complex (Fig. \ref{fig:cc} (b)).
Accordingly, the reduced barrier for hole capture facilitates carrier recombination \textit{without} optical excitation.
The subsequent electron capture process will be fast due to the negligible energy barrier (see Fig. \ref{fig:cc} (b)).
The \ce{V_S}-\ce{V_{Cu}} complex shows similar behavior, but its concentration is low under standard growth conditions.

\textit{Activation by passivation:} 
It has been suggested that donor-acceptor complexes passivate deep donors in kesterite CZTS \cite{Chen:2010jd} and chalcopyrite CIGS, \cite{Zhang:1998hz}
which make them more tolerant to defects.
However, we show that the neutral donor \ce{V^{0}_{S}}, which is deactivated by double Sn reduction,
can be reactivated by forming complexes with the ionized acceptor \ce{Cu^{1-}_{Zn}} and thus become an efficient recombination center.
This is partially because the dominant defect-defect interaction is the classical Coulomb attraction instead of quantum mechanical level repulsion as is often considered. \cite{Walsh:2017ko}

We also examine recombination pathways via the donor levels of \ce{Sn_{Zn}}.
Fig. \ref{fig:charge} (c) shows the defect charge density of \ce{Sn_{Zn}} is well localized around the Sn lone pair, 
suggesting the transitions involving Sn reduction and oxidation could trigger the carrier recombination similar to those in \ce{V_S} and \VSCZ{}.
The recombination path involving the double donor level \ce{Sn_{Zn}}(2+/1+) has a relatively high electron capture barrier of 0.23 eV (Fig. \ref{fig:cc} (c) and Table \ref{table:rate}).
On the other hand, \ce{Sn_{Zn}}(1+/0) -- corresponding to the transition between Sn(III) and Sn(II) -- has a smaller energy barrier of 0.05 eV, implying a faster recombination process.

\begin{figure*}
\centering
\includegraphics[width=12.8cm]{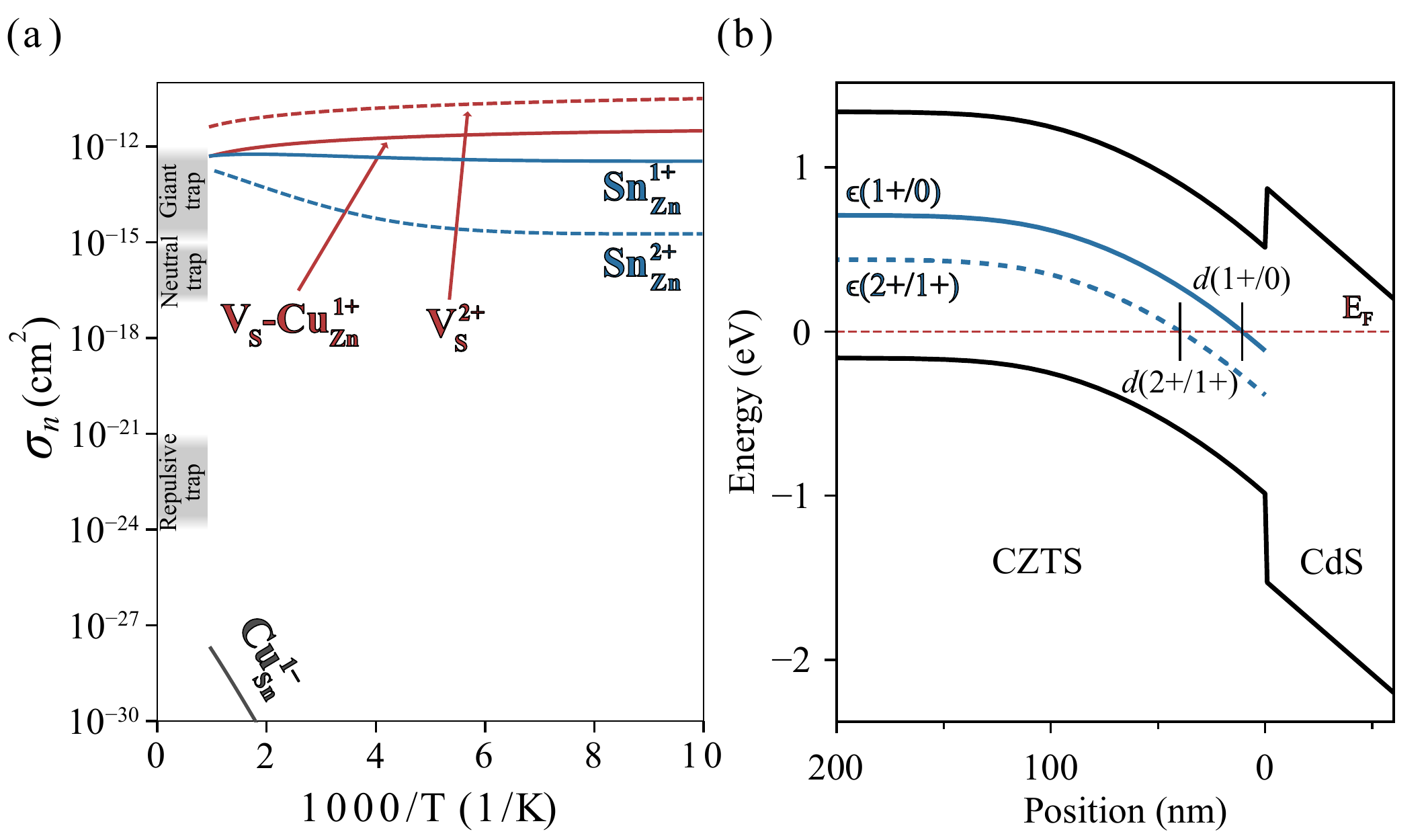}
\caption{\label{fig:rate}
(a) Electron capture cross-sections of \ce{V_S}, \VSCZ{} complex, \ce{Sn_{Zn}}, and \ce{Cu_{Sn}}.
Gray shades represent the typical orders of magnitude of cross sections of giant, neutral and repulsive traps. \cite{Stoneham:gb}
(b) Band diagram of CZTS/CdS heterojunction.
The solid and dashed blue curves represent the single $\epsilon (1+/0)$ and double donor levels $\epsilon (2+/1+)$ of \ce{Sn_{Zn}}, respectively.
$d(q_1/q_2)$ is defined by the distance from the interface where the charge transition levels of \ce{Sn_{Zn}} $\epsilon(q_1/q_2)$ equals the Fermi level ${E_F}$ (red dashed line).
The band diagram and $d(2+/1+)=40$ nm are obtained by solving the  Poisson-Boltzmann equation using \url{https://pythonhosted.org/eq_band_diagram}.
}
\end{figure*}

In Fig. \ref{fig:rate} (a), we present the capture cross section calculated within the static coupling approximation. \cite{Alkauskas:2014kk}
\ce{V_S}-\ce{V_{Cu}} and \ce{Sn_{Zn}} can be classified as a giant electron trap whose electron capture cross section ($\sim 10^{4}$ \AA$^2$) far exceeds the size of its atomic structure. \cite{Stoneham:gb}
The calculated capture cross sections of the \textit{native} defects in CZTS are orders of magnitudes larger than \textit{extrinsic} transition metal impurities in silicon solar cells including Ti, V, Cr, Mo, Fe, Au and Zn whose cross sections range from $10^{-1}$ \AA$^2$ to $10^{3}$ \AA$^2$. \cite{Macdonald:2004eh,Peaker:2012hd}
This analysis suggests that \VSCZ{} and \ce{Sn_{Zn}} are the main sources of non-radiative recombination that limit the efficiency of CZTS solar cells (see Table \ref{table:rate}). 
Note that due to the small energy barrier of \VSCZ{}, the recombination is expected to be fast even at low temperature.
At high temperature, the slight decrease in the capture cross section is attributed to the high Landau-Zener velocity; \cite{Zener:1932iz}
the faster the defect level crosses the conduction bands, the less likely the defect captures electrons.
The calculated capture cross section of \ce{Sn^{1+}_{Zn}} is an order of magnitude higher than that of \ce{Sn^{2+}_{Zn}} (Fig. \ref{fig:rate} and Table \ref{table:rate}).

In an operating solar cell, the recombination rate due to \ce{Sn_{Zn}} may depend on the spatial position
because the distance from the interface between CZTS and CdS determines the Fermi level (electronic band bending) and, hence, the charge state of \ce{Sn_{Zn}}.
In the undepleted region ($d>d(2+/1+)$ in Fig. \ref{fig:rate} (b)),
most of \ce{Sn_{Zn}} is in the form of +2 charge state which is a slower recombination channel.
However, in the depletion region ($d<d(2+/1+)$),
\ce{Sn_{Zn}} favors a +1 charge state which has much larger capture cross section.
In this case, a recombination pathway is activated by band bending in a photovoltaic device. 

\subsection*{S-rich growth environment}
Under S-rich conditions, the formation of \ce{V_S} and \VSCZ{} is strongly suppressed (see Fig. \ref{fig:form} (b)).
We associate this with the experimentally observed increase in \ce{V_{OC}} under a high S partial pressure during the annealing of a 
photovoltaic device. \cite{Ren:2017hva}
However, even under S-rich conditions, a considerable concentration of \ce{Sn_{Zn}} is still expected, 
which can limit the lifetime of carriers to below 7.1 ns (see Table \ref{table:rate}).
This shows good agreement with the reported photoluminescence (PL) decay times of kesterite materials 
which range from 1 ns to 10 ns. \cite{Hages:2017co,Yan:2018dw}
Moreover, the electron paramagnetic resonance (EPR) signal in CZTS \cite{Chory:2010fm} supports the existence partially oxidized Sn(III) with an unpaired electron ($5s^1$), which is the active state in the proposed recombination pathways.

\begin{table*}[]
\centering
\caption{\label{table:rate}
Equilibrium point defect concentrations (\ce{N_{T}}), carrier capture cross section at 330K ($\delta_{n/p}$) for electrons ($n$) and holes ($p$), thermal activation energy ($E_t$) and capture barrier ($E_b$)
of \VSCZ{} (0/+), \ce{Sn^{2+}_{Zn}} (2+/1+), \ce{Sn^{1+}_{Zn}} (1+/0), and \ce{Cu_{Sn}} (2$-$/1$-$) charge transitions.
The thermal velocities of electron and hole are 2.9$\times10^{7}$ and 1.9$\times10^{7}$ cm/s, respectively.
The Shockley-Read-Hall coefficient ($\ce{A}=R/\Delta n$) is calculated for an excess carrier concentration $\Delta n = \num{1e14} $ \ce{cm^{-3}}
}
\label{Killer Centers}
\begin{tabular}{ l l l l l l l l l l l l l} 
\hline\\[-1em]
Defect & \multicolumn{2}{c}{\ce{N_{T}} (\ce{cm^{-3}})} && \multicolumn{2}{c}{$\delta$ (\ce{cm^{2}})} & $E_t$ (eV) &\multicolumn{2}{c}{$E_b$ (meV)}  && \multicolumn{2}{c}{\ce{A} (s$^{-1}$)}&\\ 
\cline{2-3} \cline{5-6} \cline{8-9} \cline{11-12}
& S-poor & S-rich && $n$ & $p$ & & $n$ & $p$ && S-poor & S-rich \\
\hline
\\[-1em]
\VSCZ{}          ($+$/0) & \num{2.7e17} & \num{2.0e12} && \num{1.5e-12} & \num{8.4e-14} &-0.04 & 9    & 190 && \num{1.8e9}   & \num{1.3e4}  \\ 
\ce{Sn_{Zn}} (2$+$/1$+$) & \num{1.1e18} & \num{3.3e14} && \num{1.5e-14} & \num{3.2e-15} & 0.60 & 230  & 5   && \num{4.8e11}  & \num{1.4e8}  \\ 
\ce{Sn_{Zn}}    (1$+$/0) & \num{1.8e12} & \num{6.7e8}  && \num{5.5e-13} & \num{5.5e-15} & 0.87 & 54   & 184 && \num{2.7e7}   & \num{1.0e4}  \\ 
\ce{Cu_{Sn}} ($2-$/$1-$) & \num{2.7e8}  & \num{6.7e11} && \num{2.6e-34} & \num{1.3e-16} & 0.31 & 1693 & 427 && \num{2.0e-18} & \num{4.9e-15}\\ 
\hline
\end{tabular}
\end{table*}

\subsection*{Inert-pair effect}
The heavy post-transition metals (elements in groups 13, 14, 15 and 16) often exhibit oxidation states two less than the group valency,
referred to as the inert-pair effect.
The inert-pair effect is explained by the insufficient screening by $d^{10}$ electrons resulting in the $s^{2}$ lone-pair electrons tightly bound to the ion. \cite{Sidgwick:1940ul,Gillespie:1957jp,ShimoniLivny:1998bc,Walsh:2011hj}
However, the role of the inert-pair effect on the properties of defects in semiconductors has not been fully explored.

\textit{Deep defect nature:} 
We find that the inert-pair effect of Sn makes deep defects, consistent with the previous theoretical studies. \cite{Biswas:2010jf,Han:2013bg}
The ability of Sn to accommodate excess charges stabilizes the neutral state over the ionized state.
In a mixed valence compound, such as CZTS, 
the variation of Madelung potential between cation sites with formal +1, +2 and +4 oxidation states 
promotes the reduction in the valence, and the ionization is suppressed more.

\textit{Large lattice distortion:}
Sn also produces defects with large distortions during carrier capture.
Electron addition or removal from \ce{V_S}, \VSCZ{} and \ce{Sn_{Zn}} are followed by the oxidation or reduction of Sn
and are therefore accompanied by large structure change.
We find, in the Sn-related defects in CZTS, large lattice distortion quantified by Huang-Rhys factor S$\gg$1. \cite{Huang:1950fg}
Especially, in \ce{V_S} and \VSCZ{}, a radiative transition pathway is impossible due to 
the very large lattice distortion where the minimum of the excited state (Sn(IV)) is located outside of the potential energy surface of the ground-state (Sn(III)). \cite{Dexter:1955bf}

Thus, the inert-pair effect in Sn is responsible for both the deep charge transition levels and the large lattice distortion of \ce{V_S}, \VSCZ{} and \ce{Sn_{Zn}}
which make them efficient non-radiative recombination centers.
Similarly, we find the deep acceptor level of \ce{Cu_{Sn}} owing to the oxidation of Cu.
However, the electron capture rates by \ce{Cu_{Sn}} (1$-$/2$-$) are low (Table. \ref{table:rate}).
The multivalency of Cu is due to the change in the occupation of $3d$ orbital:
from $3d^{10}$ in Cu(I) to $3d^{9}$ in Cu(II).
Thus, the local relaxation after the oxidation is not significant. 
The small lattice distortion in \ce{Cu_{Sn}} produces an electron capture barrier above 1.6 eV, making capture unlikely (Fig. \ref{fig:rate} (a)).

Emergence of deep defects induced by the formation of lone pairs has also been reported in CIGS.
Extrinsic dopants of Bi and Sb are deep in \ce{CuInSe_2} due to the lone-pair $s^2$ states. \cite{Park:2014iw}
Han {et al.}\cite{Han:2016ht} have also found that the formation of lone pairs in amorphous oxide semiconductors, such as \ce{InGaZnO_4} and \ce{ZnSnO_3}, is responsible for electron trapping.

While Sn reduction captures electrons, in lone-pair compounds whose cations, such as Tl(I), Pb(II) and Bi(III), have occupied $s^2$ in the stoichiometric structure, the oxidation of cations could capture holes.
Several EPR measurements show that Pb(III) is responsible for the hole traps in lead halides.
\cite{Iwanaga:2002fv,Shkrob:2014gt,Cortecchia:2017ku,Colella:2018cj}

The efficiency and lifetime of optoelectronic devices can be severely damaged by a defect with fast non-radiative recombination.
A. M. Stoneham \cite{Stoneham:gb} suggests several characteristics of such  \textit{killer} centers, including:
(1) defects producing many and closely spaced electronic levels;
(2) defects with large lattice distortions.
The first type can be directly related to the transition metal impurities with partially filled $d$ orbitals (e.g. Ni in GaP).
While a simple vacancy center was suggested as a candidate for the second-type, \cite{Stoneham:gb}
a wide variety of vacancies are not recombination centers in photovoltaic materials (e.g. \ce{V_{Cu}} is a shallow acceptor in CZTS, and \ce{V_I} is a shallow donor in \ce{CH3NH3PbI3}.)
Because of the strong interaction between impurities and host materials, 
it is difficult to find a general trend of the properties of defects in the absence of detailed calculations.
On the other hand, we find that the inert lone-pair of Sn is the origin of the large cross-section of an wide range of defects
and not significantly altered by a specific configuration or an electronic state of the defect.
Thus, we speculate that the inert-pair effect could likely cause \textit{killer} centers with the ability to act as giant carrier traps in a broad range of semiconductors.

Many photovolatic materials with band gaps close to the theoretical optimum of 1.3--1.5 eV \cite{Shockley:1961co}
show poor performances; in particular a low \ce{V_{OC}}.
The record efficiency of \ce{Cu2SnS3}, whose band gap is around 1 eV, 
is still low (4.63\%) even with the high current density \ce{J_{SC}} of $37.3$ mA/cm$^2$
mainly because of low \ce{V_{OC}} of 283 mV. \cite{Nakashima:2015js}
The carrier lifetime of \ce{Cu2SnS3} was reported to be very short (0.1--10 ps). \cite{Baranowski:2015hd}
\ce{Sb2Se3} solar cells also exhibit low \ce{V_{OC}} with short carrier lifetime of 1.3 ns. \cite{Wen:2018fr}
The first light-to-electricity conversion efficiency of \ce{Cu3BiS3} solar cell (0.17\%) has been achieved only recently with \ce{V_{OC}} of 190 meV, \cite{Li:2018fl}
and, to our knowledge, a successful operation of \ce{CuBiS2} solar cell has not been reported.
The presence of lone-pair cations is a common feature in these technologies. 

\section*{Conclusion}
The lone-pair effect associated with Sn is responsible for both the deep defect levels 
and large lattice distortions that facilitate rapid electron-hole recombination in the semiconductor \ce{Cu2ZnSnS4}.
By employing a first-principles approach to predict the defect levels, concentrations, and capture rates,
we can distinguish between active and inactive defect centres.
For a material grown under S-poor conditions, 
\ce{V_S}, \VSCZ{} and \ce{Sn_{Zn}} act as dominant recombination centers,
while \ce{Sn_{Zn}} limits the minority carrier lifetime under S-rich conditions.
We propose that a similar mechanism could responsible for the low
performance of other emerging photovoltaic compounds. 
We emphasise the need for further experimental and theoretical investigation of defects in
semiconductors composed of heavy post-transition metals 
to further evaluate the \textit{general} role of the inert-pair effect on the non-radiative 
electron-hole recombination process.

\begin{acknowledgments}
We acknowledge support from the Royal Society, the EPSRC (grant no. EP/K016288/1), and the EU Horizon2020 Framework (STARCELL, grant no. 720907). 
We are grateful to the UK Materials and Molecular Modeling Hub for computational resources, which is partially funded by EPSRC (EP/P020194/1).
Via our membership of the UK's HEC Materials Chemistry Consortium, which is funded by EPSRC (EP/L000202), this work used the ARCHER UK National Supercomputing Service (http://www.archer.ac.uk).
\end{acknowledgments}

\bibliography{bib} 

\end{document}